\documentclass[letterpaper,twocolumn,10pt]{article}
\usepackage[available,functional, reproduced]{usenixbadges}
\usepackage{usenix}
\usepackage[normalem]{ulem}
\newcommand*{\Mname}{MGG}
\usepackage{amsmath,amsthm,mathtools}

\newcommand\fix[1]{\textcolor{blue}{#1}}

\usepackage[normalem]{ulem}

\usepackage{xcolor}
\usepackage{color,soul}
\usepackage{cancel}

\definecolor{codegreen}{rgb}{0,0.6,0}
\definecolor{codegray}{rgb}{0.5,0.5,0.5}
\definecolor{codepurple}{rgb}{0.58,0,0.82}
\definecolor{backcolour}{rgb}{0.95,0.95,0.92}
\definecolor{textblue}{rgb}{.2,.2,.7}
\definecolor{textred}{rgb}{0.54,0,0}
\definecolor{textgreen}{rgb}{0,0.43,0}
\definecolor{codered}{rgb}{201,72,12}

\newcommand\redsout{\bgroup\markoverwith{\textcolor{red}{\rule[0.5ex]{2pt}{1pt}}}\ULon}

\usepackage[normalem]{ulem}
\usepackage[T1]{fontenc}
\usepackage[scaled=0.85]{beramono} 
\usepackage{listings}
\usepackage{multirow}
\usepackage{multicol}
\usepackage{url}
\usepackage{subfig}
\usepackage{array}
\newcommand{\PreserveBackslash}[1]{\let\temp=\\#1\let\\=\temp}

\usepackage[ruled,vlined,linesnumbered]{algorithm2e}
\SetCommentSty{mycommfont}

\definecolor{calpolypomonagreen}{rgb}{0.12, 0.3, 0.17}
\definecolor{cobalt}{rgb}{0.0, 0.28, 0.67}
\lstdefinestyle{tt1}{
language=C,
basicstyle=\linespread{0.98}\ttfamily\footnotesize,
breaklines=true,
numbers=left,
frame=single,
numberstyle=\tiny, 
stepnumber=1,
numbersep=5pt, 
tabsize=4,
commentstyle=\color{textred},   
keywordstyle=\bfseries\color{codegreen},
stringstyle=\color{textgreen},
columns=fullflexible,
keepspaces=true,
xleftmargin=\parindent,
showstringspaces=false,
otherkeywords = {True, False},
keywordstyle=[3]\color{codegreen}\bfseries,
keywords=[3]{__device__},
keywordstyle=[4]\color{cobalt}\bfseries,
keywords=[4]{nvshmem_TYPE_get, nvshmem_TYPE_get_warp, nvshmem_TYPE_get_block, nvshmem_malloc, nvshmem_barrier_all, nvshmem_free, nvshmem_finalize, nvshmem_init, nvshmem_team_my_pe, cudaSetDevice, cudaMalloc, nvshmem_float_get, nvshmem_float_get_warp, nvshmem_float_get_block,
nvshmem_float_get_,
warp,
block
},
}

\lstdefinestyle{tt3}{
language=C,
basicstyle=\linespread{0.98}\ttfamily\footnotesize,
breaklines=true,
numbers=left,
frame=single,
numberstyle=\tiny, 
stepnumber=1,
numbersep=5pt, 
tabsize=4,
keywordstyle=\bfseries\color{codegreen},
commentstyle=\color{textred},   
stringstyle=\color{textgreen},
columns=fullflexible,
keepspaces=true,
xleftmargin=\parindent,
showstringspaces=false,
otherkeywords = {True, False},
keywordstyle=[2]\color{cobalt}\bfseries,
keywords=[2]{MGG_kernel, MGG_host},
keywordstyle=[3]\color{codegreen}\bfseries,
keywords=[3]{__shared__, __global__},
}
\usepackage{multirow}
\usepackage{tikz}
\usepackage{xcolor}
\usepackage{ctable} 

\newcommand*\circled[1]{\tikz[baseline=(char.base)]{
            \node[circle,fill=.,inner sep=0.8pt] (char) {\textcolor{white}{#1}};}}

\begin{document}

\date{}
\pagestyle{empty}

\title{\Large \bf MGG: Accelerating Graph Neural Networks with Fine-Grained Intra-Kernel Communication-Computation Pipelining on Multi-GPU Platforms}

\author{
{\rm Yuke Wang, Boyuan Feng, Zheng Wang, $^\dagger$Tong Geng, $^*$Kevin Barker, $^*$Ang Li, and Yufei Ding}\\
$^\dagger$University of Rochester, $^*$Pacific Northwest National Laboratory\\
University of California, Santa Barbara
} 

\maketitle

\begin{abstract}
The increasing size of input graphs for graph neural networks (GNNs) highlights the demand for using multi-GPU platforms.
However, existing multi-GPU GNN systems optimize the computation and communication individually based on the conventional practice of scaling dense DNNs. For irregularly sparse and fine-grained GNN workloads, such solutions miss the opportunity to jointly schedule/optimize the computation and communication operations for high-performance delivery. 

To this end, we propose \textbf{MGG}
\footnote{Paper is accepted to OSDI'23.}
, a novel system design to accelerate full-graph GNNs 
on multi-GPU platforms.
The core of MGG is its novel dynamic software pipeline to facilitate fine-grained computation-communication overlapping within a GPU kernel.
Specifically, MGG introduces GNN-tailored pipeline construction and GPU-aware pipeline mapping to facilitate workload balancing and operation overlapping.
%
%
{MGG also incorporates an intelligent runtime design with analytical modeling and optimization heuristics to dynamically improve the execution performance.}
Extensive evaluation reveals that MGG outperforms state-of-the-art full-graph GNN systems across various settings: on average $4.41\times$, $4.81\times$, and $10.83\times$ faster than DGL, MGG-UVM, and ROC, respectively.
%
\end{abstract}
\section{Introduction}
Over the recent years, graph-based deep learning has attracted lots of attention from the research and industry communities. 
Among various graph-learning methods, graph neural network (GNN)~\cite{GCNConv, GINConv, GATConv} gets highlighted most due to its success in many deep learning tasks (\textit{e.g.}, node feature vector (embedding) generation for node classification~\cite{kaspar2010graph, gibert2012graph, duran2017learning} and link prediction~\cite{chen2005link, kunegis2009learning, tylenda2009towards}).
GNNs consist of several layers, where layer $k+1$ computes the embedding for a node $v$ based on the embeddings at the previous layer $k$ ($k \geq 0$) by applying 
\begin{gather*} \small \label{eq: GNN}
 \begin{aligned} 
  a_{v}^{(k+1)}  &= \mathbf{Aggregate}^{(k+1)}({h_{u}^{(k)}|u\in \mathbf{N}(v)\cup h_v^{(k)}})  \\
  h_{v}^{(k+1)}  &= \mathbf{Update}^{(k+1)}(a_{v}^{(k+1)})
\end{aligned}   
\end{gather*}
where $h_{v}^{(k)}$ is the embedding of node $v$ at layer $k$.
The $\mathit{Aggregate}$ function accumulates neighbors'($\mathbf{N}(v)$) embeddings of node $v$. The $\mathit{Update}$ function consists of a fully-connected NN layer. The neighbor aggregation ($\mathit{Aggregate}$) is the key bottleneck that dominates the overall computation due to its high computation sparsity and irregularity~\cite{GNNAdvisor, HyGCN}.
Compared with conventional graph analytics (e.g., random walk~\cite{grover2016node2vec, deepWalk}), GNN features higher accuracy~\cite{GCNConv, GINConv} and better generality~\cite{SageConv, zhang2020every} on various applications.

GNN computation on large input graphs (millions/billions of nodes and edges) usually counts on powerful multi-GPU platforms (e.g., NVIDIA DGX~\cite{dgx-a100}) for {scaling up the performance}. 
{
The multi-GPU system (that can potentially store all data required for the computation in the aggregate memory of all GPUs on a single machine)}
can benefit from aggregated memory capacity and bandwidth (HBM and NVLinks) with more GPUs. There is also a popular trend for state-of-the-art hyper-scale systems employing GPU-centric building blocks. For example, the recent NVIDIA DGX SuperPod~\cite{dgx-superpod} consists of 32$\times$DGX-H100 servers (each with 8$\times$H100).
Unfortunately, the runtime performance of GNNs does not scale proportionally with the aggregated compute capability and memory capacity of the platform. 
{This is mainly because the irregular and sparse local memory access of neighbor aggregation in the single-GPU settings now ``scales'' to more expensive inter-GPU communication (i.e., remote memory access).}
Such intensive inter-GPU communication becomes the new critical path of multi-GPU GNN execution and offsets the performance gains from multi-GPU computation parallelism. 

Based on this observation, we highlight a more promising way of formalizing GNN computation on multi-GPU systems. Our key insight is that GNN execution can be more precisely abstracted as a fine-grained dynamic software pipeline to encourage communication and computation overlapping, which will largely hide the communication cost.
The opportunities for building such fine-grained pipelines widely exist at different granularities in GNNs. 
For instance, on a single graph node, the remote neighbor access can be overlapped with the local neighbor computation. Among  different graph nodes, the remote neighbor access for certain nodes would potentially be overlapped with the local neighbor computation of some other nodes.
However, prior research could hardly exploit such benefits since they rely on hardware and software infrastructures tailored for coarse-grained~\cite{sysMLROC, PyTorch-Direct} and regular communication patterns~\cite{ma2019neugraph, P3}. 
To capitalize on the fine-grained pipelining benefits, there are three major challenges. 

The first challenge is \textit{how to craft the pipeline structure}. A work-efficient pipeline for GNNs demands comprehensively considering multiple factors (e.g., the operations and the number/granularity of each pipeline stage) to best fit the GNN algorithm and multi-GPU computation/communication.
The second challenge is \textit{how to map the pipeline to the GPU processing units}. 
Given the GPU's architectural complexity (e.g., multi-granular processing units and multi-layer memory hierarchy), different mapping and primitive choices would bring performance and design flexibility tradeoffs. 
The third challenge is \textit{how to find and adapt toward the ``optimal'' pipeline configuration swiftly}. 
Given the diversity of GNN inputs (\textit{e.g.}, graph structures) and hardware (\textit{e.g.}, different types/numbers of GPUs), pinpointing the best-off design configuration with high-performance delivery relies on combined insights from the properties of the software pipeline, GNN inputs, and GPU programming and execution paradigms. 

\begin{figure} [t] \small
\centering\includegraphics[width=\columnwidth]{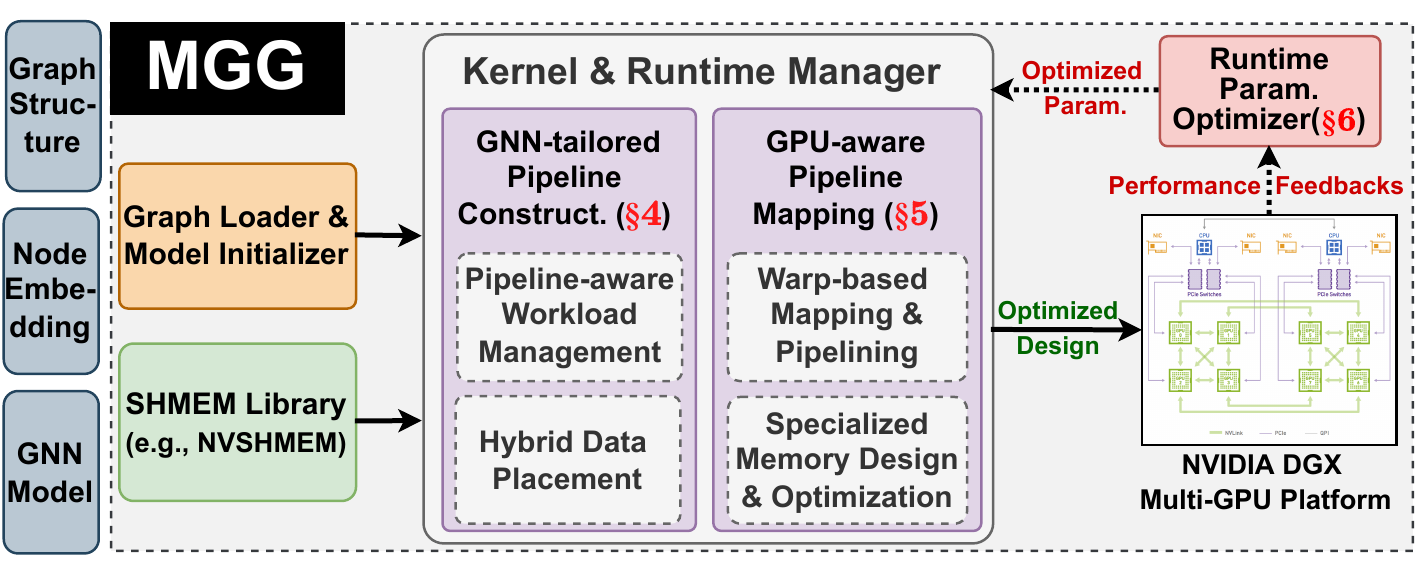}
    \vspace{-18pt}
    \caption{Overview of \Mname.}
    \label{fig: Overview}
    \vspace{-15pt}
\end{figure}
To this end, we introduce a set of principles for multi-GPU GNN acceleration via a fine-grained dynamic software pipeline.
\textit{To construct fine-grained pipelines}, the original coarse-grained irregular GNN computation should be breakdown into fine-grained operations.
The joint optimization of the GNN workload granularity and data layout should be carried out to facilitate operation overlapping.
%
%
\textit{To map pipelines to GPUs}, the proper GPU logical processing units (e.g., thread, warp, and block) should be selected for promoting GPU kernel efficiency and design flexibility. 
In addition, the right choice of communication primitives (e.g., NVSHMEM~\cite{nvshmem}) should be determined to provide fine-grained inter-GPU communication support.
{\textit{To adapt pipelines dynamically}}, customized kernel templates with tunning knobs should be devised. This will help to maintain pipelining effectiveness across a diverse range of GNN inputs and hardware platform settings.

We crystallize the above principles into \Mname\footnote{\url{https://github.com/YukeWang96/MGG-OSDI23-AE.git}}, a holistic system design and implementation for multi-GPU GNNs (Figure~\ref{fig: Overview}).
Given the GNN models and inputs, \Mname~will automatically generate pipeline-centric GPU kernels for multi-GPU platforms and dynamically improve the kernel performance based on runtime feedback.
%
%
The core of \Mname~is its \textbf{Kernel \& Runtime Manager}, which constructs GNN-tailored pipelines and maps such pipelines to proper  communication primitives and GPU logical processing units. 
It can also dynamically orchestrate GPU kernels based on new configurations.
\Mname~also incorporates a \textbf{Runtime Parameter Optimizer}, which will monitor the performance (e.g., latency) from the actual execution and generate new configurations for the next iteration based on the analytical performance model and optimization heuristics.
%
To the best of our knowledge, we are the first to explore the potential of GPU kernel operation pipelining for accelerating irregular GNN workloads. Moreover, \Mname~can be generalized to other applications (e.g., deep-learning recommendation model (DLRM)~\cite{naumov2019deep}) that are sharing similar irregular communication demands (\fix{$\S$\ref{sect: additional study}}). 

Overall, we make the following contributions in this paper:
\begin{itemize}
    \vspace{-3pt}
    \item We propose a GNN-tailored pipeline construction technique ($\S$\ref{sect: GNN-tailored Pipeline Construction}) with pipeline-aware workload management and hybrid data placement, for efficient communication-computation pipelining in a GPU kernel.
    \vspace{-3pt}
    \item We introduce a GPU-aware pipeline mapping strategy ($\S$\ref{sect: GPU-aware Pipeline Mapping}), encompassing warp-based mapping and pipelining, and specialized memory designs and optimizations to comprehensively promote kernel performance.
    \vspace{-3pt}
    \item We devise an intelligent runtime with lightweight analytical modeling and optimization heuristics to dynamically improve the performance of GNN training
    ($\S$\ref{sect: Modeling and Design Optimization}). 
    \vspace{-3pt}
    \item Comprehensive experiments demonstrate that MGG can outperform state-of-the-art multi-GPU GNN systems across various GNN benchmarks. Additionally, MGG can be generalized to other DL applications, like DLRM.
\end{itemize}

\section{Related Work}
\label{sect: Background and Motivation}
Recent deep-learning applications expand their scope from handling structured dense inputs (e.g., images) to unstructured sparse inputs (e.g., graphs).
Along with such algorithmic/application expansion is the exploration of new system designs and optimizations for more efficient deep learning.
One of the most important topics is the ability to handle large-scale inputs, which are usually out of the computation and memory capacity of one GPU. 
For scaling regular deep-learning applications, like dense DNNs, various abstractions (e.g., data and model parallel) and high-performance communication libraries (e.g., NCCL~\cite{nccl}) have been developed. While the scaling approach for irregular GNN applications is still initial and suffers from unsatisfactory performance. 

Compared to scaling dense DNNs, scaling sparse GNNs is significantly more challenging. 
The irregular fine-grained sparse GNNs workload cannot fit the regular coarse-grained workload abstraction for dense DNNs. 
The cost of irregular communication in GNNs cannot be easily amortized by simply batching more requests as dense DNNs due to their randomness and sparseness. 
Scaling strategies largely vary among different GNN inputs while tiling/schedule strategies would be reused across different inputs of dense DNNs. 
{Therefore, an array of dedicated designs have been introduced to scale the sparse GNNs, focusing on three major directions.}
\begin{figure} [t] \small
    \centering
\includegraphics[width=0.9\linewidth]{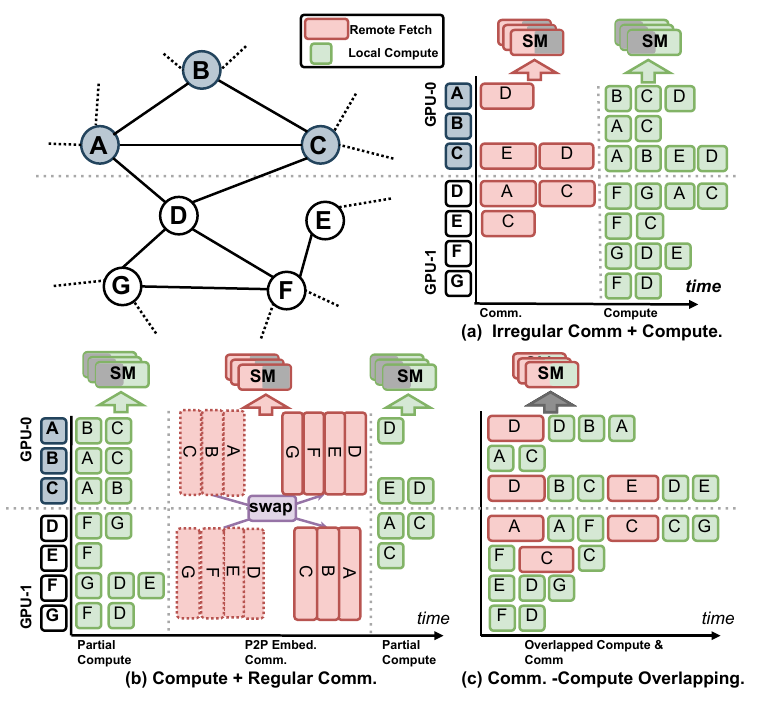}
    \caption{Different Multi-GPU GNN strategies for computation and communication. Note that red and green boxes indicate aggregation workload on remote and local neighbors. ``SM'' boxes with grey areas indicate potential idleness.} 
    \label{fig: comparison with different Multi-GPU solution.}
    \vspace{-10pt}
\end{figure}

\textbf{Operator Specialization for Sparse Communication:} This is the mainstream solution that treats the communication as a standalone operator for irregularly sparse GNN communication (Figure~\ref{fig: comparison with different Multi-GPU solution.}(a)). 
DGL~\cite{wang2019dgl} is the state-of-the-art GNN framework and its most recent update incorporates PyTorch-Direct~\cite{PyTorch-Direct} (a GNN-tailored communication design based on zero-copy memory~\cite{zero-copy}) for large-scale GNN training across GPUs.
Work from~\cite{cai2021dgcl} introduces a communication planning algorithm for distributed GNNs by considering links, communication, contention, and load balancing.
However, these efforts optimize the communication standalone and thus miss the opportunities to jointly optimize computation and communication operations/schedules which can potentially reduce the overall latency and improve GPU utilization. 

\textbf{Algorithm Modification for no Communication:} 
The second typical type is to eliminate irregular communication by altering algorithms~\cite{pagraph,yang2022gnnlab,wang2019dgl}. They harness various algorithmic adaption solutions, such as neighbor sampling and mini-batch to prefetch the remote neighbors to local devices, and then train the GNN model in a data-parallel fashion as the traditional dense DNN. 
However, existing research~\cite{sysMLROC, chen2018fastgcn} shows that such an algorithmic modification would compromise the accuracy of GNN models compared to the original GNNs. It would also destabilize the algorithmic performance (\textit{e.g.}, the lower convergence speed and final accuracy) under different inputs and sampling configurations.

\textbf{Schedule Transformation for Dense Communication: }
The third type is to transform irregular communication to regularized communication (e.g., AlltoAll, P2P), which has been optimized by existing communication kernels (Figure~\ref{fig: comparison with different Multi-GPU solution.}(b)). 
ROC~\cite{sysMLROC} delegates communication to its underlying NVIDIA Legion runtime~\cite{bauer2012legion}, which manages irregular remote neighbor access via a DMA engine. It batches fine-grained embeddings into large embedding tiles on CPUs to facilitate coarse-grained data movement between the host and GPUs.
NeuGraph~\cite{ma2019neugraph} tiles the large node embedding matrices by rows (as embedding chunks) and then forwards each chunk to GPUs sequentially via coarse-grained P2P communication. 
%
P3~\cite{P3} spots the potential of transforming irregular embedding communication to regular all-to-all communication for embedding column tiles.
However, this type of effort would introduce many unnecessary data movements and non-trivial overhead to transform original algorithms and data inputs.  

To sum up, existing designs explore solutions in a limited scope and have yet to extend their solution search to a broader context by exploring the synergy between the multi-GPU GNN workloads, GPU execution paradigms, and communication patterns. Therefore, these designs could hardly enjoy the full potential of multi-GPU platforms.

\section{Motivation}
Different from prior solutions, we propose a new view for multi-GPU GNN workload. 
We spot that by removing the explicit barrier between the computation and communication stage in multi-GPU GNNs, we can co-schedule the operations from both stages in a holistic way that can reduce the GPU resource idleness and promote performance (Figure~\ref{fig: comparison with different Multi-GPU solution.}(c)). 
For example, when GPUs initiate remote access requests and are waiting for the arrival of remote data, the idle cycles of GPUs can be fulfilled by other local computing workloads.
Such insight enables us to abstract the multi-GPU GNN workload as a fine-grained dynamic software pipeline for communication and communication overlapping. 
{Specifically, ``Fine-grained'' {means that} the operations at each pipeline stage are tiny {(e.g., the aggregation of one neighbor's embeddings)} versus DNN layers.``Dynamic'' {means} that {the division of computation into pipeline stages} would vary among different inputs in contrast to DNNs with a relatively fixed pipeline.
}
Such a new design is motivated by our three major observations. 

\textbf{GNN Workload Speciality:} 
{The first observation reveals the specialty of GNN workloads, which feature two major types of partial dependency that facilitate pipelining~\cite{alle2013runtime}.}
The first type is the fine-grained neighbor aggregation dependency, where the neighbor embeddings of individual graph nodes are aggregated either sequentially or in parallel with proper synchronization. 
The second type is the dynamic execution dependency on limited processing units, where different operations would compete for limited GPU resources (e.g., SMs) during the runtime.
Such two types of dependencies expose new opportunities for us to amortize communication costs by overlapping neighbor aggregation from different nodes.
\begin{figure}[t]
\begin{lstlisting}[style=tt1, caption={NVSHMEM APIs in CUDA C.}, label={code: wmma API interface.}]
// Initialize an NVSHMEM context on CPUs.
nvshmem_init();
// Get the current GPU device ID on CPUs.
int gpu_id = nvshmem_team_my_pe(NVSHMEMX_TEAM_NODE);
// Set the GPU based on its device ID on CPUs.
cudaSetDevice(gpu_id);
// Define NVSHMEM memory visible for all GPUs on CPUs.
d_shared_mem = (void*) nvshmem_malloc (num_bytes); 
// Define global memory visible only for the current GPU.
cudaMalloc((void**) &d_mem, num_bytes); 
// Remote access API called by a thread/warp/block.
__device__ nvshmem_float_get_{warp/block}(void *dst, const void *src, size_t nelems, int src_gpu_id);
// Sync all GPUs within an NVSHMEM context on CPUs.
nvshmem_barrier_all();
// Release NVSHMEM objects on CPUs.
nvshmem_free(d_shared_mem);
// Terminate the current NVSHMEM context on CPUs.
nvshmem_finalize();
\end{lstlisting} 
\vspace{-6pt}
\end{figure}

\textbf{GPU Execution Characteristics:} 
{The second observation highlights the characteristics of the GPU execution paradigm.} One key design principle of GPUs is their massive computation/communication parallelism to amortize the unit cost of individual computation/communication operations~\cite{ryoo2008optimization}. The underlying mechanism of GPU hardware design to facilitate this is to simultaneously schedule multiple logical processing units (e.g., threads/warps/blocks) to share the hardware processing units (i.e., GPU SMs). 
Such a design provides the essential ingredient for pipelining, which is that computation and communication operations can co-run on the same units at the same time to fulfill the idle GPU cycles and maximize the utilization of the GPU hardware processing units. 
Moreover, with the precise control of GPU kernel launching parameters (e.g., the size of the block and shared memory), the effectiveness of co-running heterogeneous operations can be adjusted so that we can  flexibly accommodate different inputs while maintaining high-performance delivery.

\textbf{Multi-GPU Programming Support:} 
The third observation features the recent advancement of the GPU communication technique and its programming support.
The one highlighted most is the NVSHMEM~\cite{nvshmem}, which provides GPU intra-kernel APIs for fine-grained (several to tens of bytes) inter-GPU communication (Listing~\ref{code: wmma API interface.}). 
{NVSHMEM is {the main} communication backend for MGG. 
Other existing techniques such as Zero-copy memory can also serve as an alternative to NVSHMEM for fine-grained communication. The performance will be similar while NVSHMEM offers better programmability.}
Some other traditional strategies for inter-GPU communication, would either offer too coarse-grained communication solutions (e.g., unified virtual memory~\cite{unifiedMemory} uses KB-level communication granularity) or resort to the default communication strategies of existing multi-GPU-based runtime system (e.g., NVIDIA Legion~\cite{bauer2012legion}) without GNN-tailored communication optimization.
\begin{figure*} [t] \small
    \centering\includegraphics[width=1\linewidth]{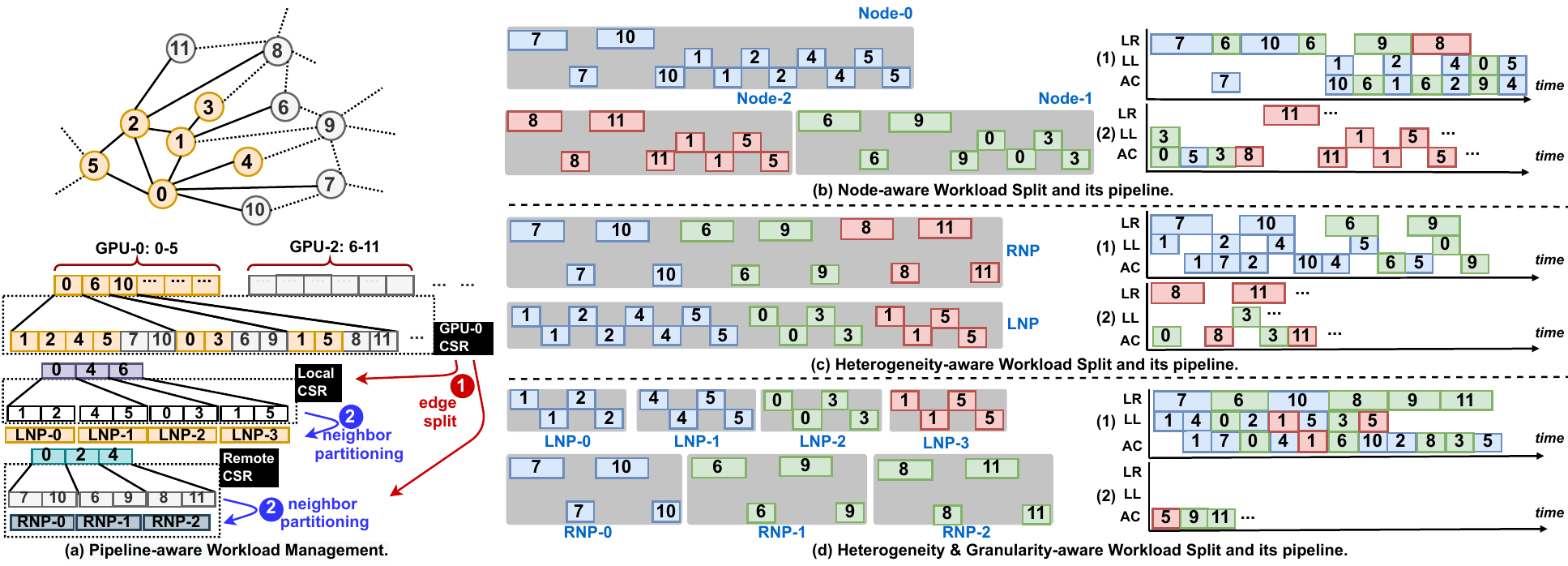}
    \caption{(a) Pipeline-aware workload management. ``LNP''/``RNP'' indicate local/remote workload partitions. (b)(c)(d) Different strategies of workload decomposition and pipelining. Each box indicates a certain (local/remote) aggregation workload and its length indicates its relative latency. ``LR'': loading remote neighbors, ``LL'': loading local neighbors, ``AC'': aggregation computation. Each grey rectangular shadow indicates a workload partition to be processed by one GPU processing unit. (1) and (2) indicate that the same pipeline is chunked into two parts along its time axis due to space limitations.} 
    \label{fig: Pipeline-aware Workload Management}
    \vspace{-5pt}
\end{figure*}

These observations and insights motivate \Mname, a holistic multi-GPU GNN system with a novel view of GNN workloads as an operation pipeline. \Mname~automates the pipeline construction, detailed pipeline mapping, and dynamic input-driven pipeline adaption, to improve the GNN scaling.

\section{GNN-tailored Pipeline Construction} 
\label{sect: GNN-tailored Pipeline Construction}
Constructing a GNN-tailored pipeline are facing two major challenges:
\textit{1) How to effectively partition and schedule multi-GPU GNN workloads} so that pipeline efficiency can be maximized;
{\textit{2) How to properly layout input} so that the hierarchy of GNN inputs and the memory/storage of multi-GPU systems can be carefully matched to facilitate pipeline execution.}
\Mname~addresses these challenges with  \textit{Pipeline-aware Workload Management} and \textit{Hybrid GNN Data Placement}.
\setlength{\textfloatsep}{6pt}

\subsection{Pipeline-aware Workload Management}
\label{sect: Pipeline-aware Workload Management}
Managing irregularly sparse GNN workloads for pipelining is challenging and could hardly benefit from the prior practice and exploration of the DNN pipeline~\cite{narayanan2019pipedream, narayanan2021memory}.

\textbf{Difference from DNN pipeline} 
\underline{\textit{First}}, balancing the GNN workloads among GPUs has to jointly optimize the computation capacity and the computation/communication irregularity. While the DNN pipeline only needs to balance the computation/memory capacity, since its pipeline stages are well-structured and their inputs are regularly dense. 
{Distributed DNNs require dense regular communication (e.g., Allreduce) that is naturally fit for existing GPU interconnects optimized for throughput and {has been} optimized by many libraries (e.g., NCCL). In contrast, distributed full-graph GNN {(with the entire graph cached on GPUs)} is much more challenging since it requires sparse irregular communication that is naturally at odds with the existing hardware interconnects, and fewer efforts have optimized its performance.}
\underline{\textit{Second}}, the GNN pipeline workload is more irregular and non-structural and can easily cause pipeline stalls/bubbles. For example, remote neighbor aggregation would have different stages (remote access + aggregation) compared with local neighbor aggregation (local access + aggregation), making it challenging to mix those two heterogeneous workloads.
While in the DNN pipeline, all inputs should consistently pass through the same pipeline stages.
\underline{\textit{Third}}, GNN pipeline stages are more fine-grained (e.g., fetching individual embeddings) compared with coarse-grained layers (e.g., GEMMs and Convolutions) in the DNN pipeline. Such small workload granularity enables different pipeline stages to overlap with each other on GPU processing units, like Streaming Multiprocessors (SMs). In contrast, DNN pipelines can only overlap layer-wise computation and communication operations among different GPUs. 

{With the above insights, we propose a three-stage dynamic software pipeline design. \fix{}
The three stages include \textit{loading remote neighbors} (\textbf{LR}), \textit{loading local neighbors} (\textbf{LL}), and \textit{aggregation computation} (\textbf{AC}). 
Aggregation of a certain neighbor will only take two stages. 
The remote neighbor aggregation will take the stage LR and AC while local neighbor aggregation will take the stage LL and AC. 
The stage-wise pipelining is achieved with two steps: 1) assigning aggregation workload to different GPU logical processing units (LPUs), like warps and blocks, and 2) scheduling different LPUs on the same GPU SM to overlap their execution. 
{Three-phase pipeline can generalize to different GNN models, which essentially consist of the different numbers of basic remote and local operations. For example, GCN has a lower local-vs-remote operation ratio while GAT features a higher local-versus-remote operation ratio.
Three-phase pipeline can also capture differences among inputs. For instance, a more sparse graph will have a higher remote-to-local operation ratio.}
}

{However, the direct construction and execution of such three-stage pipelines would be inefficient, because of its ignorance of GNN workload heterogeneity and irregularity on multi-GPU platforms.
To address these challenges, \Mname~highlights a GNN-tailored pipeline construction strategy to build and optimize the software pipeline in three steps.} 
%
%

\textbf{Step-1: Workload-aware inter-GPU pipeline workload balancing.} 
{This step aims to construct the ``raw'' pipeline and balance workloads among pipelines on different GPUs. 
Our insight is that GPUs with massive processing units (e.g., SMs) will serve many pipelines concurrently, and the key to maximizing GPU performance and utilization is to ensure that each pipeline will get a similar amount of workload, thereby avoiding execution critical path on certain ``long'' pipelines.} 
We, therefore, develop a \textit{range-constrained binary search algorithm} (Algorithm~\ref{algo: Range-constrained Binary Search.}) based on prior graph partitioning exploration~\cite{andreev2004balanced}. Our solution features a lower runtime cost to split the GNN input graph into chunks (one chunk per GPU) while balancing the number of edges within each chunk. 
%
%
%
Then the workload from the same chunk is grouped by nodes as \textit{workload partitions} mixed local and remote neighbors (Figure~\ref{fig: Pipeline-aware Workload Management}(b)). 
From its potential execution pipeline, we can see many idle cycles (indicated by blank spaces in different pipeline stages) which would result in low pipeline efficiency and GPU resource occupancy. 
Note that in the software pipeline, workloads from different partitions can be overlapped as they will be processed by different LPUs. While the workloads from the same partition are sequentially processed by one LPU and their relative order should be maintained even after being mixed with other partitions. 

\textbf{Step-2: Heterogeneity-aware pipeline bubble reduction.} 
{The pipeline constructed from the previous step is still inefficient due to its scattered workloads among stages, namely pipeline bubbles.
The optimization in this step is to minimize such pipeline bubbles for better pipeline efficiency.
The key is to reduce the heterogeneity of workload partitions that hinders effective overlapping.}
%
To achieve this, we categorize the sparse multi-GPU GNN computation into two types. The first type has local neighbor access only, which has shorter execution latency. The second type has remote neighbor access, which features high latency overhead. We delicately handle different types of workloads via grouping (Figure~\ref{fig: Pipeline-aware Workload Management}(a)-\circled{1}), where two separate CSRs for \textit{local} and \textit{remote} subgraphs will be built. The aggregation will be conducted on local and remote subgraphs separately and followed by a result synchronization at the end. Such a remote-local split is also backed by the fact that on platforms with all-to-all GPU interconnections (e.g, DGX-A100/H100), accessing different GPUs under the same data granularity {has approximately equal communication cost~\cite{li2019evaluating}}. 
Such heterogeneity awareness in workload partitioning (Figure~\ref{fig: Pipeline-aware Workload Management}(c)) enables a more densely overlapped workload between the stage LR and LL/AC.

\textbf{Step-3: Granularity-aware intra-GPU pipeline enhancement.} 
{While the second optimization improves pipeline efficiency by reducing the workload heterogeneity, there is still plenty of room for further enhancement. 
The optimization in this step is to facilitate a more balanced workload distribution among pipeline stages.
This key is to find the proper workload granularity for local and remote subgraphs so that those originally sequentially processed workload partitions can be overlapped.}
Our key observation is that nodes in the local/remote subgraphs would have a diverse number of neighbors. 
Such a specialty makes it challenging for massively parallel GPUs to harvest the real performance gains due to the imbalance workload and diverged execution flow. 
\begin{algorithm}[t] \small
  \caption{Range-constrained Binary Search.}
  \label{algo: Range-constrained Binary Search.}
\SetAlgoLined
  \SetKwInOut{Input}{input}
  \SetKwInOut{Output}{output}
  \SetKwFunction{FMain}{binSearch}
  \SetKwProg{Fn}{Function}{:}{}

  \Input{Graph node pointer array ($\mathit{nPtr}$), edge list array ($\mathit{eList}$) , and the number of GPUs ($\mathit{numGPUs}$).}
  \Output{list of graph edge split points ($\mathit{numGPUs} - 1$).}
    $\mathit{outList} = \{\}$\; $\mathit{lastPos} = 0$\;
    \tcc{Compute approximated \#edges per GPU.}
    $\mathit{ePerGPU}$ = $(len(eList) + \mathit{numGPUs} - 1)/\mathit{numGPUs}$\;
    \For{$\mathit{sId}$ \textbf{in} [0, 1, ..., $\mathit{numGPUs} - 1$]}{
        $nid$ = $\mathit{\texttt{binSearch}}(\mathit{nPtr}, \mathit{ePerGPU}, \mathit{lastPos}, \mathit{numNodes})$\;
        $\mathit{lastPos} = nid$\;
        $\mathit{outList[sId]} = nid$\;
    }
    \textbf{return} $outList$\;
\BlankLine
\tcc{Search split points on $nPtr$.}
\Fn{\FMain{$nPtr$, $ePerGPU$, $lastPos$,                      $numNodes$}}{
      $i = lastPos$\; 
      $j = numNodes$\;
      $target = \min(nPtr[i] + ePerGPU, nPtr[numNodes])$\;
        \While{$i < j$}{
        $mid = (nPtr[i] + nPtr[j])/2$\;
        \uIf{$mid > target$}{    
            $j = (i + j) / 2$\;
        }
        \uElse{
            $i = (i + j) / 2$\;
        }
        }
    \KwRet i\;
 }
\end{algorithm}
Therefore, we approximate such coarse-grained irregular workloads with fine-grained fixed-sized partitions so that the workload imbalance across nodes can be amortized. 
For example, with 2 neighbors per partition (Figure~\ref{fig: Pipeline-aware Workload Management}(a)-\circled{2}), we can get a more balanced workload among nodes in their local and remote neighbor aggregation. 
With such granularity awareness, the individual pipeline can be further condensed along its time axis with more overlapping of the LL and AC stage. (Figure~\ref{fig: Pipeline-aware Workload Management}(d)). 
Meanwhile, the irregular workload can be more evenly distributed to GPU SMs for higher GPU utilization.
On the other side, partition granularity should also be balanced with synchronization overhead, since more fine-grained partitioning can bring more parallelism at the cost of more synchronization overhead. This is because workloads from different partitions for the same target node need to be reduced via synchronization, like inter-thread shuffling and atomics.


\Mname~design can also be generalized to multiple machines with a minor adaptation.
For example, in Figure~\ref{fig: Pipeline-aware Workload Management}(d), when there are inter-node (over Inifite-Band) remote neighbors (longer latency due to lower inter-node communication speed), the size of remote neighbor partitioning (RNP) should be adjusted to a smaller size (e.g., from 2 to 1 remote neighbor) to facilitate better overlapping with local computation.

\subsection{Hybrid GNN Data Placement} \label{sect: Hybrid GNN Data Placement}
In collaboration with our multi-step pipeline construction, 
we introduce a \textit{hybrid GNN data placement} strategy  to exploit the benefits of different types of memory in SHMEM-enabled multi-GPU systems. 
{The major impact of such hybrid placement on pipelining is two-fold. 
First, placing GNN data in different memory spaces will lead to different ratios of local and remote workloads, thus, affecting workload balance among pipelines. 
Second, different memory spaces will offer different access performances (e.g., latency), thereby, affecting the execution efficiency of the individual pipelines, such as the number of pipeline bubbles.
}

\begin{figure} [t] \small
    \centering
    \includegraphics[width=\linewidth]{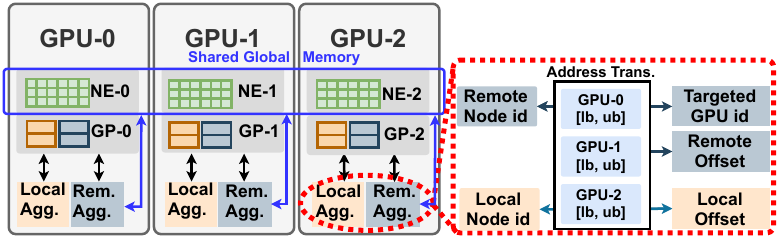}
    \caption{\Mname~Storage Layout and Communication Pattern. Note that ``NE-i'' is the node embedding partition stored on the i-th GPU. ``GP-i'' is the neighbor partition processed by the i-th GPU. ``GPU-i [lb, ub]'' is the node-id range [lowerbound, upperbound] of the node embeddings on the i-th GPU.}
    \label{fig: NVSHMEM Design and Communication Pattern.}
    \vspace{5pt}
\end{figure}

Our strategy focuses on two major aspects. Firstly, for workload balance among pipelines, we leverage NVSHMEM ``shared'' global memory to store the node embeddings (\textbf{NEs}) of the whole graph (Figure~\ref{fig: NVSHMEM Design and Communication Pattern.} \textit{left}). Our major  consideration here is that such shared global memory space can be accessed by all GPUs with the approximated equal access speed, which is vital to facilitate a more even distribution of remote workloads to GPUs in terms of their size and unit access costs. 
In addition, NEs are generally large in terms of size (due to high dimensionality), which are beyond the device memory limit of a single GPU. 
Therefore, NEs are ideal to be placed in shared global memory space with sufficient space (with aggregated memory of different GPUs), which also provides direct remote access support across GPUs.
Specifically, we will partition the NEs of input graphs into $n$ equal-sized partitions (where $n$ is the number of GPUs) and place each of them in one GPU's shared global memory space.

{Secondly, for the efficiency of individual pipelines,} we allocate the ``private'' global memory space for storing partitioned graph structure (\textbf{GP}) data, which is only visible to kernels on the current GPU. 
{Our key insight is that GP (e.g., edge lists), is all scalar values and usually small in size, and will only be accessed by the local GPU. Therefore, GP is ideal to be placed in individual GPUs' DRAM. Such a placement is also important to reduce unnecessary and inefficient remote access on those tiny scalars for fewer pipeline bubbles.}
In our design, GP data (e.g., edges) from private GPU global memory will be processed by a \textit{address translation} unit for fetching correct NEs on local/remote GPU since the NE indices are rebased to zero on each GPU (Figure~\ref{fig: NVSHMEM Design and Communication Pattern.} \textit{right}).

\section{GPU-aware Pipeline Mapping}
\label{sect: GPU-aware Pipeline Mapping}
Efficient pipelining also demands effective mapping of well-constructed pipeline workload and their schedules to the low-level GPU logical processing units (\textit{e.g.}, GPU threads/warps/blocks) 
to overlap computation and communication. To achieve this, we propose \textit{Warp-based Mapping \& Pipelining}  and \textit{Specialized Memory Design \& Optimization} to jointly optimize the pipeline execution efficiency, GPU utilization, and end-to-end design flexibility.

\subsection{Warp-based Mapping \& Pipelining}
\label{sect: Workload Interleaving and Warp-based Mapping}
{An effective pipeline mapping demands comprehensive consideration of two major aspects. 
1) \textit{Which type of GPU logical processing units (e.g., warps, blocks) should be used for pipeline workload partitions?} }
We choose GPU \textit{warp} as the basic working unit to handle the workload of each partition. 
This is because threads in a warp can collaboratively work on different dimensions of a node embedding simultaneously. Whereas using a single or several threads (less than the size of a warp, 32 threads) would hardly explore the computation parallelism and would cause warp-level divergence. 
Besides, NVSHMEM remote access initiated by a warp of threads would merge the requests into one remote memory transaction to amortize the overhead.
\begin{figure} [t] \small
    \centering
    \includegraphics[width=\linewidth]{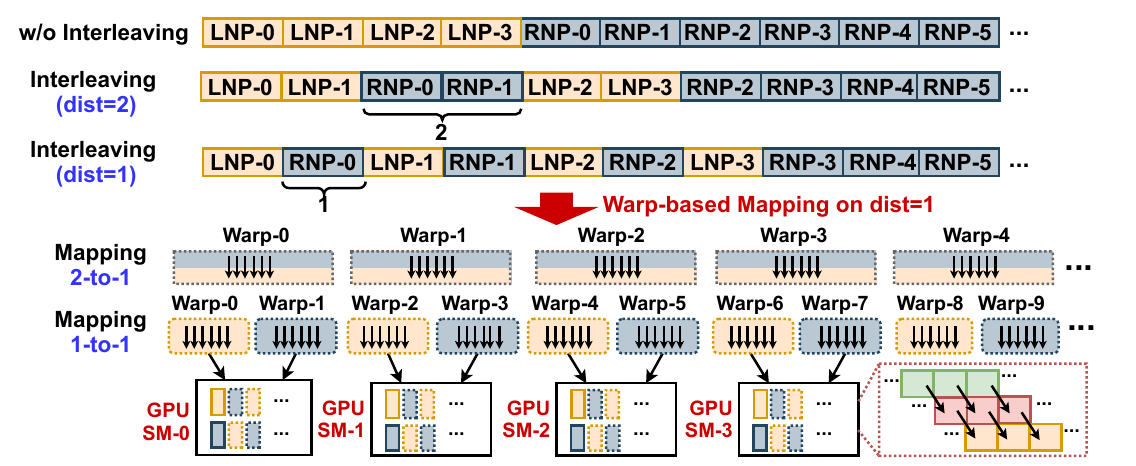}
    \caption{Warp-based Mapping and Pipelining. Note that ``LNP'' refers to the local neighbor partitions; ``RNP'' refers to the remote neighbor partitions. Workload and Warps are matched based on colors. Tiny boxes in GPU SM indicate decomposed workload operations for overlapped execution.}
    \label{fig: Workload Interleaving and Warp-based Mapping.}
    \vspace{8pt}
\end{figure}
{2) \textit{Which pattern of mapping should be used for benefiting pipeline execution efficiency?}} 
The most straightforward way is to continuously map the neighbor partitions from the local and remote workload list to GPU warps with continuous IDs (Figure~\ref{fig: Workload Interleaving and Warp-based Mapping.}).
However, this strategy would easily suffer from workload imbalance among GPU SMs. 
This is because warps with continuous IDs are more likely to be placed into the same thread block, which is assigned to one SM for processing. Therefore, SMs assigned with warps for handling remote neighbor partitions would lead to much longer latency than SMs assigned with warps for processing local neighbor partitions. Such a workload imbalance would lead to poor GPU utilization and runtime execution performance. 

To this end, we introduce our novel \textit{workload interleaving} strategy to balance the workload among SMs on GPUs. 
Each warp of threads running on GPU would handle one or more pairs of local/remote workload partitions. 
To more precisely calibrate the warp-to-SM mapping for different pipeline stages to achieve efficient pipelining, we introduce a new metric -- \textit{interleaving distance}. 
We give examples with the interleaving distance equals 1 and 2 for illustration (Figure~\ref{fig: Workload Interleaving and Warp-based Mapping.}).
{By mixing different types (both local and remote) of workload together, better GPU utilization can be achieved since when one warp is blocked for high-cost remote access, other warps that are working on local computation can still be served by the SMs warp scheduler for filling up these idle GPU cycles.} 
Moreover, such a design would improve design flexibility. 
For instance, given an input graph with a selected neighbor partition size, we can adjust the size of interleaving distance and the workload per warp so that waiting cycles of the remote access can be hidden by the computation cycles of the neighbor aggregation. 
Thus, each warp can be fully utilized while the design can achieve sufficient parallelism.

{MGG currently processes the neighbors of adjacent nodes (based on node-ids) to the same thread block where the same block will be scheduled on the same SM. If there are common remote neighbors for those adjacent nodes, their remote requests will be merged.
Improving such locality requires reordering the graph nodes to maximize their common neighbors. Such an exploration is orthogonal to our current contribution.
In future GPUs, there is a trend to explore the locality among independent processing units. For instance, in Hopper, several thread blocks can be grouped together as thread-block groups. We can explore the tradeoff between the locality benefits and group synchronization overhead.}

\subsection{Specialized Memory Design \& Optim.} 
\label{sect: Specialized Memory Design & Optim.}
Efficient software pipelining also demands careful management of high-bandwidth shared memory for promoting data access efficiency and asynchronized primitives for exploiting intra-warp operation pipelining.

\textbf{GPU SM Shared Memory Layout:} Based on our MGG's warp-based workload design, we propose a \textit{block-level shared memory orchestration} to maximize the performance gains. We have several key insights for such a dedicated memory layout design within each thread block. 
\underline{\textit{First}}, our neighbor-partition-based workload will generate the intermediate results that can be cached at the high-speed shared memory for reducing the frequent low-speed global memory access.
\underline{\textit{Second}}, NVSHMEM-based remote data access demands a local scratch-pad memory (e.g., registers, shared and global memory) to hold the remote data for local operations. 

For the \textit{local} neighbor aggregation, we reserve a shared memory space with $\mathit{D}$ ($D$ is the embedding dimension) floating-point numbers for embeddings of the target node in each neighbor partition so that threads from a warp can cache the intermediate results of partial reduction in shared memory. 
For the \textit{remote} neighbor aggregation, the shared memory space is doubled $\mathit{2\times \mathit{wpb} \times \mathit{D}}$ ($\mathit{wpb}$ is the warps per block). The reason is that we need the first half $\mathit{wpb} \times D$ for caching the partial aggregation results of each warp and the remaining for the remotely accessed neighbor embeddings. 
For each MGG kernel design, we will first identify the warp-level information, like warp IDs. Then within each thread block, we define the customized shared memory layout by splitting the contiguous shared memory address into three different parts for neighbor ids, partial aggregation results, and the remotely-fetched node embeddings. 
%
%
We use the dynamic shared memory for design flexibility since those parameters (\textit{e.g.}, \textit{wpb} and \textit{D}) can only be determined at runtime. 
During execution, we will first calculate the total shared memory size per block and then pass it as a kernel launching parameter. 

\textbf{Pipelined Memory Operation:}
$\S$\ref{sect: Workload Interleaving and Warp-based Mapping} have discussed assigning local (LNP) and remote (RNP) neighbor aggregation workloads to warps so that different warps can overlap their computation and communication to fully saturate the active cycles of the GPU SM scheduler. However, only exploiting the inter-warp communication-computation overlap is not enough to maximize the utilization of GPU resources. We further explore the overlapping of the computation and communication at the intra-warp level by carefully scheduling the memory operations. 
\begin{figure} [t] \small
    \centering
    \includegraphics[width=\linewidth]{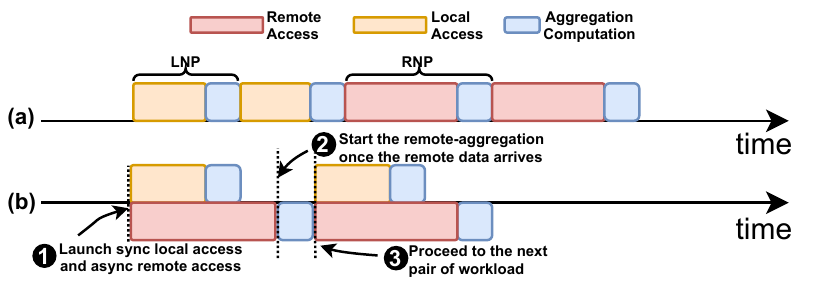}
    \caption{Illustration of (a) w/o and (b) w/ asynchronized primitives for overlapping computation and communication of an individual warp. Note that the length of each rectangular box indicates the estimated latency cost of each operation.}
    \label{fig: pipeline memory operations}
    \vspace{10pt}
\end{figure}
Figure~\ref{fig: pipeline memory operations}(a) shows the case with two LNPs and two RNPs by using the synchronized remote access, we can just sequentially process the two LNPs and the two RNPs. The long-latency remote access can happen only after the completion of its preceding LNP. This could lead to a longer GPU stall for memory operations and low GPU SM utilization. 
Our profiling also shows that without overlapping, the remote access usually dominates the overall execution (around 60\% of overall latency) compared to the time for local data access plus the time for aggregation computation (around 40\% of overall latency). Such observation justifies our design to mainly hide the latency from remote access.

To amortize the cost of remote access for each warp, we introduce \textit{asynchronized remote memory operations} (Figure~\ref{fig: pipeline memory operations}(b)). This improved design consists of two major steps. First, we can simultaneously launch the local memory access while initializing the remote memory access for fetching the node embedding (\circled{1}), therefore, the time for remote access can be amortized by the processing of LNP. Second, once the remote access is completed, the current warp will start aggregation on the remotely-fetched node embedding data (\circled{2}). The next step will start the new iteration of the previous two steps, which will process a new pair of LNP and RNP. 
\section{Intelligent Runtime Design}
\label{sect: Modeling and Design Optimization}
In this section, we will discuss our intelligent runtime design with performance/resource analytical modeling and heuristic-based cross-iteration optimization strategy.

\textbf{Performance-Resource Analytical Modeling:}
The performance/resource model of \Mname~has two variables: \textit{workload per warp} ($\mathit{WPW}$) and \textit{shared memory usage per block}
($\mathit{SMEM}$), which can be measured by
\begin{equation} \small
\label{equ: workload per threads}
\begin{split}
    \mathit{WPW} &= 2\cdot\mathit{ps}\cdot{D}\cdot\mathit{dist}, \\
    \mathit{SMEM} &=\mathit{ps}\cdot\mathit{wpb}\cdot{IntS} + 2\cdot\mathit{wpb}\cdot{D}\cdot\mathit{FloatS}
\end{split}
\end{equation}
where $\mathit{ps}$, $\mathit{wpb}$, and $D$ are the sizes of neighbor partition, warp per block, and node embedding dimension, respectively; 
$\mathit{dist}$ is the interleaved distance of local/remote workloads ($\S$\ref{sect: Workload Interleaving and Warp-based Mapping});
$\mathit{IntS}$ and $\mathit{FloatS}$ are both 4 bytes on GPUs.
To determine the value of the $\mathit{ps}$, $\mathit{wpb}$, and $\mathit{dist}$ of a given input graph, we will first 
compute the total number of warps by using  
\begin{equation}  \small
\label{equ: numWarps}
\begin{split}
    \mathit{numWarps} &= \frac{\max\{local, remote\}}{dist}
\end{split}
\end{equation}
where $\mathit{local}$ and $\mathit{remote}$ are the number of local and remote partitions, respectively.
Then we compute the total number of blocks and the estimated block per SMs by using 
\begin{equation}  \small
\label{equ: numBlocks}
\begin{split}
    \mathit{numBlocks} &= \frac{\mathit{numWarps}}{\mathit{wpb}}, \\
    \mathit{blocksPerSM} &= \frac{\mathit{numBlocks}}{\mathit{numSMs}}
\end{split}
\end{equation}
\vspace{3pt}

Later, based on our micro-benchmarking results on diverse datasets, we define our parameter search space and constraints: 
1) $ps\in[1\dots32]$ to balance the computation parallelism and synchronization overhead;
2) $\mathit{dist}\in[1\dots16]$ to effectively overlap the computation and remote memory access;
3) $\mathit{wpb}\in [1\dots16]$ to maintain SM warp scheduling flexibility for better occupancy and throughput; 
4) $numSMs\leq{c_1}$, $SMEM\leq{c_2}$, where $c_1$ and $c_2$ are hardware constraints~\cite{GPU-micro-arch}, \textit{e.g.}, NVIDIA A100 has 108 SMs and 164KB shared memory per SM.

\textbf{Heuristic-based Cross Iteration Optimization}
To optimize the design of \Mname, the parameter $\mathit{ps}$, $\mathit{dist}$, and {{\textit{wpb}}} are initialized as the value 1 at the beginning. Then we optimize one parameter in each of the following iterations. 
{\textit{First}}, we increase the $\mathit{ps}$ to maximize the warp utilization. When further increasing the $ps$ would also increase the latency, we would stop the search on $\mathit{ps}$ and switch to $dist$.
{\textit{Second}}, we apply a similar strategy to locate the value of $\mathit{dist}$ that can maximize the overlap of local computation and remote access.
{\textit{Third}}, we increase $\mathit{wbp}$ to maximize the utilization of the entire SM. If any increase of $\mathit{wpb}$ would increase the latency, we know that there may be too large thread blocks or too heavy workloads on individual warps that lower SM warp scheduling efficiency or computation parallelism. 
We would ``retreat'' (\textit{i.e.}, decrease) $\mathit{ps}$ to its second-highest value if necessary and restart the increase of $\mathit{wpb}$.
This optimization algorithm will stop when any decrease of $\mathit{ps}$ and increase of $\mathit{wpb}$ would lead to higher latency than the top-3 lowest latency.
The latency of each iteration during the optimization will be recorded by a configuration lookup table. Finally, the configuration with the lowest latency will be applied.

This particular optimization order of parameters ($\mathit{ps}$, $\mathit{dist}$, and $\mathit{wpb}$) is based on two major aspects: (i) \textit{Spatially speaking}, the granularity is from coarse-grained algorithm-level partitioning through $\mathit{ps}$, to medium-grained pipeline construction through $\mathit{dist}$ (according to the partition plan), to fine-grained pipeline-to-warp fine-tuning through $\mathit{wpb}$ (according to the pipeline design).
(ii) \textit{Temporally speaking}, the three optimizations are applied at loading-time ($\mathit{ps}$ to decide layout), kernel initialization ($\mathit{dist}$ to decide pipeline), and runtime ($\mathit{wpb}$ to decide pipeline mapping), respectively.

The {above parameter adaption for dynamic pipelining is vital for design/optimization generality. This is because the characteristics of graphs (\#nodes/edges and embedding sizes) would lead to different efficiency of kernel pipelines. {Our later experimental studies (as shown in Figure~\ref{fig: Parameter Selection for Four Settings}) demonstrate} its benefits with up to 70\% of performance improvements.}
\section{Evaluation} 
\label{sect: evaluation}
\begin{table}[t] \small
\centering
\caption{Datasets for Evaluation.}
\vspace{-6pt}
\scalebox{0.75}{
 \begin{tabular}{ l | r r r r }
\specialrule{.1em}{.05em}{.05em} 
\textbf{Dataset} & \textbf{\#Vertex} & \textbf{\#Edge} & \textbf{\#Dim} & \textbf{{\#Class}}\\
\hline\hline
reddit(\textbf{RDD})~\cite{wang2019dgl}            & 232,965   & 114,615,892 & 602 & 41 \\ 
enwiki-2013(\textbf{ENWIKI})~\cite{snapnets}    & 4,203,323 & 202,623,226 & 300 & 12 \\ 
it-2004 (\textbf{IT04})~\cite{suitesparse}       & {41,291,594} & {1,150,725,437} & 256 & 64  \\ 
ogbn-paper100M(\textbf{PAPER})~\cite{P3}       & {111,059,956} & {1,615,685,872} & 128 & 64  \\ 

{ogbn-products(\textbf{PROD})}~\cite{hu2020open}    & {2,449,029} & {61,859,140} & 100 & 47 \\ 
{ogbn-proteins(\textbf{PROT})}~\cite{hu2020open}    & {132,534} & {39,561,252} & 8 & 112 \\
com-orkut(\textbf{ORKT})~\cite{snapnets}        & 3,072,441         & 117,185,083 & 128 & 32 \\ 
\specialrule{.1em}{.05em}{.05em} 
\end{tabular}}
\vspace{5pt}
\label{table: Evaluation Dataset}
\end{table}
\paragraph{Benchmarks \& Datasets}
Despite the diversity of GNN models, the fundamental computation and communication paradigm (vector-based scatter-gather operation) in multi-GPU GNNs remains the same. We evaluate two distinctive and representative GNN models on \textit{node classification} tasks: 

The first type of GNN model uses a \textit{non-discriminated} neighbor aggregation strategy, where all neighbors contribute equally when doing the aggregation. We choose \textbf{Graph Convolutional Network (GCN)}~\cite{GCNConv}, which is the most popular GNN model and is also the key backbone network for many other GNNs, such as GraphSAGE~\cite{SageConv} and Differentiable Pooling~\cite{diffpool}. 
We use \textit{2 layers with 16 hidden dimensions} for GCN, which is also the setting from the original paper~\cite{GCNConv}.
The computation of a 2-layer GCN can be expressed as 
\begin{equation} \small
    Z = \mathit{Softmax}(\hat{A}\ ReLU(\hat{A}XW^1)W^2).
\end{equation}
where $\hat{A}$ is the adjacent matrix of the input graph with self-loop edges, and $X$ is the input node embedding matrix, where $X\in R^{N\times D}$; $N$ is the number of nodes in a graph; $D$ is the size of node embedding dimensions. $W^1$ and $W^2$ are trainable weight matrices in layer-1 and layer-2, respectively. 

The second type uses a \textit{discriminated} neighbor aggregation strategy, where neighbors would contribute differently depending on their calculated edge-specific features. We choose 
\textbf{Graph Isomorphism Network (GIN)}~\cite{GINConv}, which aims to distinguish the graph structure that cannot be identified by GCN. Each layer of GIN can be expressed as
\begin{equation} \small
h_{v}^{l+1} = \mathit{MLP}^{l}((1+\epsilon^{l})\dot h^{l} + \sum_{u\in N_{(v)}}h_{u}^{l}).    
\end{equation}
where $l$ is the layer ID and $l\in\{0,1\}$, $\mathit{MLP}$ is a fully-connected neural network, $h_{v}$ is the node embedding for node $v$, and $N_{(v)}$ stands for the neighbors of node $v$. 
GIN mainly differs from GCN in its aggregation function, which introduces a weight parameter as the ratio of contribution from its neighbors and the node itself. 
In addition, GIN is the reference architecture for many other advanced GNNs with more edge properties, such as Graph Attention Network~\cite{GATConv}. For GIN evaluation, we use \textit{5 layers with 64 hidden dimensions}, which is also the setting used in the original paper~\cite{GINConv}. 
Graphs (Table~\ref{table: Evaluation Dataset}) used in our evaluation are large in their number of nodes and edges that demand multi-GPU capability for effective GNN computation. 
\textit{\#Class} is the output dimension (\textit{\#labels}) for the node classification task. \textit{\#Dim} is the embedding dimension of the input graph.

\textbf{Baselines} In this evaluation, we compared \Mname~with several existing systems that support large full-graph GNN {(i.e., caching the entire graph on GPUs)} on multi-GPU platforms.
\underline{\textbf{1) Deep Graph Library (DGL)}}~\cite{wang2019dgl} is the state-of-the-art framework for large-scale GNNs across GPUs. It leverages PyTorch-Direct~\cite{min2021pytorch} as the communication backend for GPU-initiated zero-copy memory access~\cite{zero-copy} to fetch neighbors embedding from the CPU host.
%
%
%
\underline{\textbf{2) \Mname-UVM}}~\cite{kim2020batch} is a GNN design by adapting \Mname~to leverage unified virtual memory (UVM). 
UVM has been highlighted in handling irregular graph computations (such as PageRank) on large graphs~\cite{kim2020batch}. 
However, \cite{kim2020batch} is not open-sourced, we thus generalize the pipeline kernel designs and optimizations ($\S$\ref{sect: GNN-tailored Pipeline Construction} and $\S$\ref{sect: GPU-aware Pipeline Mapping}) of \Mname~to build such a UVM baseline and incorporate optimizations from~\cite{kim2020batch}. 
{Note that UVM and zero-copy memory are different communication backends [1]. Thus, MGG-UVM does not implement zero-copy data transfer. We remark UVM is the key communication protocol before the new hardware support for fine-grained direct GPU-GPU communication (e.g., NVSHMEM). UVM is more coarse-grained and {will require the engagement of CPUs (e.g., host memory management) for communication}. The reason to use MGG-UVM is to show that {if there is no advanced hardware} support (e.g., NVSHMEM) for fine-grained direct GPU-GPU communication, the benefits of our elaborated pipeline can be offset by UVM communication overhead.}
\begin{figure}[t] \small
    \centering
     \subfloat[GCN Model.]
    {\includegraphics[width=0.95\columnwidth]{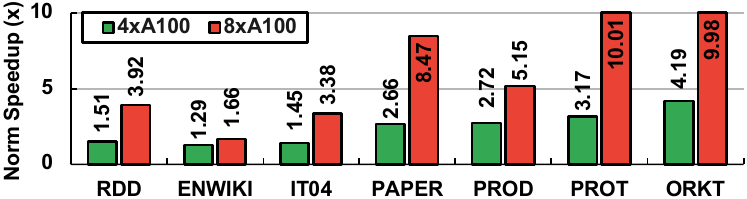}} 
     \\ 
    \subfloat[GIN Model.]
    {\includegraphics[width=0.95\columnwidth]{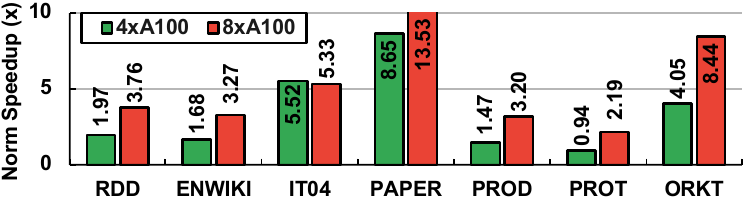}}
     \\ \vspace{-5pt}
    \caption{Performance comparison with DGL. Note that full-graph PAPER on DGL requires A100-80GB.}
    \label{fig: compare with Pytorch-direct}
    \vspace{5pt}
\end{figure}
\underline{\textbf{3) ROC}}~\cite{sysMLROC} is a popular distributed multi-GPU system for full-graph computation. ROC highlights its learning-based partitioning and leverages NVIDIA Legion~\cite{bauer2012legion} runtime for communication and task scheduling. 

Other multi-GPU GNN designs, like NeuGraph~\cite{ma2019neugraph} and $P3$~\cite{P3}, are not publicly available. Initially, we plan to evaluate \Mname~on AMD ROC\_SHMEM~\cite{rocsheme}. However, as indicated in its document, the existing ROC\_SHMEM is an experimental prototype and is not officially ready to be applied in practice due to very strict software limitations (\textit{e.g.}, only supports ROCm v4.3) and hardware (\textit{e.g.}, only supports AMD GFX9 GPUs), which are quite challenging to find and deploy and not supported by any existing GNNs frameworks~\cite{cai2021dgcl, PyTorch-Direct, sysMLROC} for comparison. We believe that once ROC\_SHMEM becomes ready and generally applicable, \Mname~can be easily migrated to AMD multi-GPU platforms.

{There is no existing design that can leverage GPU-to-GPU communication only for distributed full-graph GNN computation}. We try our best to measure the best-possible baseline performance. DGL and ROC have longer latency {in the earlier iteration} due to cache warmup for node embedding on GPU memory. We thus perform 
{warm up iterations until their per-iteration latency} becomes stable, and then measure their performance with minimized CPU-GPU data movements.

\textbf{Platforms \& Tools} 
\label{sect: Platforms and Metrics }
The implementation of \Mname~consists of $\sim$9K LoC. 
We compile and link \Mname~with CUDA (v11.2), OpenMPI (v4.1.1), NVSHMEM (v2.0.3), and cuDNN (v8.2) library.
Our major platform is an NVIDIA DGX-A100 with dual AMD Rome 7742 processors (each with 64 cores, 2.25 GHz), 1TB host memory, and 8$\times$A100 GPUs (40 GB) connected via NVSwitch, which offers 600 GB/s GPU-to-GPU bi-directional bandwidth. 
For the modeling study, we also leverage DGX-1 with 4$\times$V100 GPUs connected via NVLinks.
We use NVIDIA NSight Compute to get the  kernel-level profiling metrics.
Speedup is averaged over 100 runs.

\subsection{End-to-End Performance}
\label{sect: End-to-End Comparison}
\textbf{Compared with DGL} In this section, we will compare with the state-of-the-art DGL framework, which leverages PyTorch-Direct for cross-GPU communication. We evaluate different datasets and platform settings (with 4 and 8 A100 GPUs).
As shown in Figure~\ref{fig: compare with Pytorch-direct}, \Mname~outperforms DGL with averaged 4.25$\times$ and $4.57\times$ speedups on GCN and GIN models, respectively. 
We also notice a trend that \Mname~demonstrates a more pronounced speedup with more GPUs. With the increasing number of GPUs, DGL suffers from heavy memory access contention, since multiple GPUs are initiating massive requests to access the neighbor embeddings on the CPU host memory.
\begin{figure}[t] \small
    \centering
    \subfloat[GCN Model.]{\includegraphics[width=0.95\columnwidth]{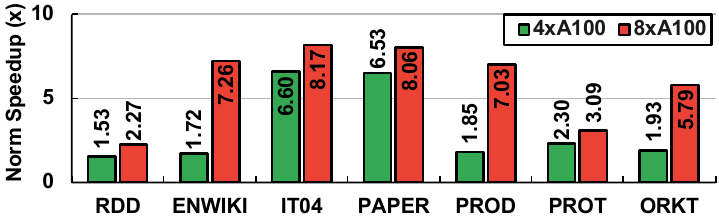}} 
    \\ 
        \subfloat[GIN Model.]{\includegraphics[width=0.95\columnwidth]{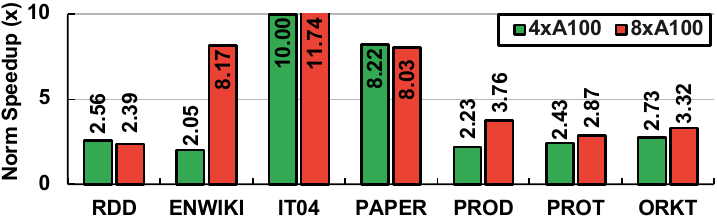} }
    \vspace{-5pt}
    \caption{Performance comparison with \Mname-UVM.}
    \label{fig: compare with UVM}
\end{figure}
Another observation is that on GIN ($D$ = 64) with higher hidden dimensionality for smaller datasets (e.g., PROD and PROT), the performance gap between DGL and \Mname~is smaller compared to GCN ($D$ = 16) since as indicated in \cite{PyTorch-Direct}, zero-copy memory would be beneficial from more coarse-grained data movement (with larger embedding vector) that can saturate the PCIe cache line (128 Bytes). While such an advantage of DGL diminishes for those larger datasets (e.g., IT04 and PAPER) on GIN due to significantly increased sparsity and irregularity. 
In addition, compared with \Mname, DGL assumes the one-size-fits-all communication strategy would work well for all input datasets. Therefore, it ignores the importance of the inputs and hardware properties, which would bring non-trivial (more than 30\%) benefits ($\S$\ref{sect: Optimization Analysis}).

{\Mname~can also be extended to cover other GNN models. The following results show the speedups of MGG over DGL on GraphSAGE with layerwise node neighbor sampling and GAT with dot-product edge attention. Table~\ref{tbl: sage and gat} shows that the performance results of GAT and SAGE also agree with our prior observations on the GCN and GIN, demonstrating the generality and effectiveness of our proposed design and optimizations to handle more complex dataflow (e.g., edge attention and softmax) in multi-GPU GNN computation.}
\begin{table}[t] \small
\centering
\caption{Additional performance comparison of MGG and DGL on GraphSAGE and GAT.}
\vspace{-5pt}
\scalebox{0.78}{
\begin{tabular}{l||r|r|r|r|r|r|r}
\specialrule{.1em}{.05em}{.05em}
\textbf{Model} & RDD & ENWIKI & IT04 & PAPER & PROD & PROT & ORKT \\
\hline
\hline
SAGE & 4.97$\times$ & 1.76$\times$ & 1.99$\times$ & 3.53$\times$ & 7.05$\times$ & 3.39$\times$ & 3.53$\times$ \\
\hline
GAT & 2.65$\times$ & 1.62$\times$ & 2.06$\times$ & 3.04$\times$ & 2.06$\times$ & 3.39$\times$ & 3.04$\times$ \\
\hline
\specialrule{.1em}{.05em}{.05em}
\end{tabular}}
\label{tbl: sage and gat}
 \end{table}

Despite that \Mname~(NVSHMEM)~and DGL (with CPU-GPU zero-copy memory~\cite{zero-copy}) both rely on GPU-initiated communication and overlap communication with computation, their underlying mechanism is different, and \Mname~shows more performance advantages. 
\Mname~can leverage inter-GPU communication while DGL can only rely on CPU-GPU communication with limited bandwidth.
This makes the communication costs pronounced in DGL and offsets the performance gains from massive thread-level parallelism.
This experiment also shows that \Mname~can serve as a drop-in replacement for the existing communication backend of DGL to improve large-scale full-graph GNN computation.

\textbf{Compared with MGG-UVM}
In this experiment, we compare MGG with its UVM-based counterpart, MGG-UVM, which uses UVM in place of NVSHMEM for remote communication. Figure~\ref{fig: compare with UVM} shows that \Mname~achieves $4.58\times$ speedup and $5.04\times$ speedup on average compared to \Mname-UVM on GCN and GIN, respectively. 
The \Mname-UVM leverages the page-faulting-based remote data access that is more coarse-grained (around 4 KB) in comparison with a single node embedding size (less than 0.4KB), which leads to higher overhead and lower effective bandwidth usage per embedding transfer. Such an overhead would exacerbate with more GPUs and also make \Mname-UVM challenging for GPU SM schedulers to effectively dispatch instructions for the next available warps. This is mainly because most of the warps wait for the long-cycle page-faulting and migration. 

We notice that with the increase of the dimension size (i.e., data movement granularity), the speedup over \Mname-UVM becomes higher. 
We later found out that the increase of data-movement granularity actually increases the overall page-fault counts. This is because embedding vectors are generally stored continuously for memory efficiency instead of aligning with the size of memory pages. Therefore, increasing the size of individual embedding also increases the likelihood of triggering multiple pagefaults per embedding transfer. 
\begin{figure}[t] \small
    \centering
\includegraphics[width=0.95\columnwidth]{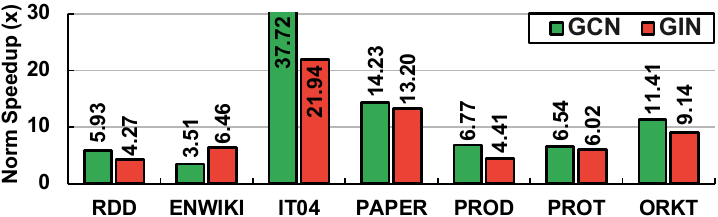}
    \caption{Performance comparison with ROC with 8$\times$A100.}
    \vspace{10pt}
    \label{fig: Comparisons with ROC.}
\end{figure}
\begin{figure*}[t] \small
    \centering
    \subfloat[]
    {\includegraphics[width=0.29\textwidth, trim=0cm -0.1cm 0 0, height=2.1cm]
    {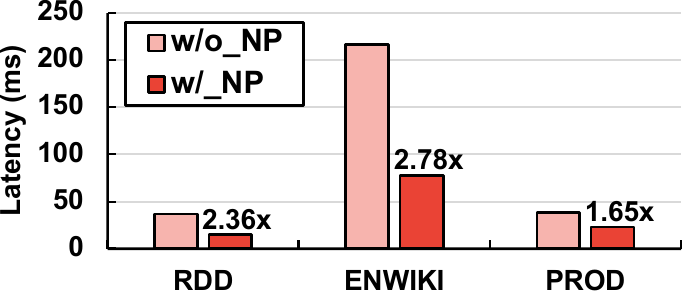}
    }
    \subfloat[]{\includegraphics[width=0.29\textwidth, trim=0cm 0cm 0 0, height=2.1cm]{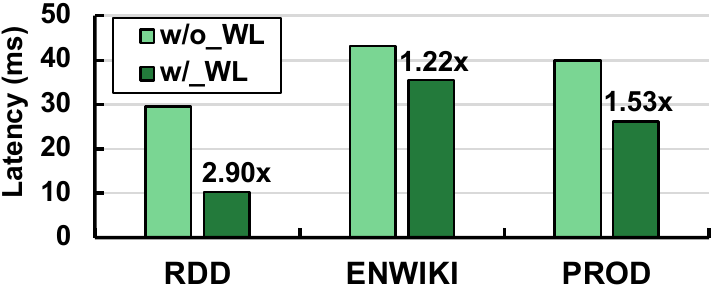}}
      \subfloat[]{\includegraphics[width=0.35\textwidth, trim=0cm -0cm 0 0, height=2.1cm]{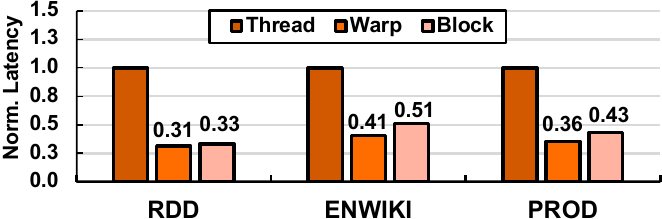}}
    \vspace{-5pt}
    \caption{Optimization Analysis: (a) Neighbor Partitioning; (b) Workload Interleaving; (c) Choice of Communication Primitives.}
    \label{fig: Optimization Analysis}
    \vspace{-5pt}
\end{figure*} 

Comparing among datasets, for graphs (e.g., PAPER) with more nodes/edges and lower average node degree, \Mname~would demonstrate more speedups since these graphs exhibit more irregular and sparse access that can not well fit into regular fix-sized pages. This also indicates the importance of amortizing communication overhead. Thanks to pipeline-centric workload management, we can effectively amortize such costs with careful operation scheduling. 

We further measure two performance-critical GPU kernel metrics that are the key indicators of our pipeline efficiency ($\S$\ref{sect: Pipeline-aware Workload Management}): \textit{Achieved Occupancy} (the ratio of the average active warps per active cycle to the maximum number of warps supported in an SM) and \textit{SM utilization} (the utilization of all available SMs on a single GPU). 
\Mname~improves SM utilization (by $21.15$\% on average) and occupancy (by $39.20\%$ on average) compared to \Mname-UVM. This indicates that \Mname~can effectively 1) distribute irregular workloads to SMs to balance workloads among pipelines and improve the overall GPU utilization, and 2) overlap the remote access and local aggregation computation from different warps to reduce pipeline bubbles and maximize SM occupancy.

\textbf{Compared with ROC}
In this experiment, we compare \Mname~with ROC~\cite{sysMLROC} on their officially released GCN model implementation. We originally plan to evaluate both 4 and 8 GPU settings. However, ROC reports many out-of-memory (OOM) errors for those large graphs on GCN/GIN model and medium graphs on the GIN model due to its aggressive caching of those intermediate tensors on GPUs. Therefore, we keep our comparison to 8 GPUs.
Performance-critical ROC runtime configurations (e.g., \#CPU cores, GPU/host memory size) are optimized to fully utilize the DGX-A100. 

Figure~\ref{fig: Comparisons with ROC.} shows that \Mname~achieves averaged $12.30\times$ and $9.35\times$ speedups over ROC on GCN and GIN, respectively.
\Mname~demonstrates a more pronounced speedup over ROC on the larger graph (e.g., IT04 and PAPER), which has more irregular neighbor embedding access.
%
The Legion runtime of ROC relies on the DMA engine for bulky data (batched embeddings) transfer between host and GPU memory, leading to higher throughput but inferior latency performance.
Besides, ROC relies on a separate communication-computation design, where computation happens after the full completion of communication. Such a design eliminates the opportunity to fill idle GPU cycles with computation during communication.
In addition, the learning-based partitioning (to reduce communication) of ROC shows benefits on relatively smaller datasets (e.g., RDD and PROT) but hard to find optimal partition plans for large graphs due to the input structure complexity. 

\subsection{Optimization Analysis}

\label{sect: Optimization Analysis}
\textbf{Neighbor Partitioning (NP)} 
We compare \Mname~with a baseline design without applying the neighbor partitioning technique (\textit{i.e.}, each aggregation workload consists of all local/remote neighbors) on 4$\times$A100. 
We apply the workload interleaving for both implementations and fix the warp-per-block size to 2 to eliminate the impact from other performance-related factors.
Figure~\ref{fig: Optimization Analysis}(a) shows higher latency (averaged $2.26\times$) for designs without applying neighbor partitioning, since the workload imbalance becomes more severe across different warps without neighbor partitioning, especially for those graphs with many remote access demands, leading to limited computing parallelism and GPU underutilization.

\textbf{Workload Interleaving (WL)}  
We compare \Mname~with a baseline design without workload interleaving (\textit{i.e.}, remote neighbor aggregation and local neighbor aggregation are mapped separately to the GPU warps. 
We fix the neighbor partition size to 16 and the warp-per-block size to 2. Figure~\ref{fig: Optimization Analysis}(b) shows that \Mname~consistently outperforms the non-interleaved baseline with an average of $1.89\times$ speedup. Without interleaving the local/remote workload, the workload distribution would be highly skewed, where the heavy and intensive remote aggregation would be gathered on certain warps close to each other while the lightweight local aggregation would be gathered on some other warps close to each other. This leads to inefficient warp scheduling and higher latency.

\begin{figure} [t] \small
    \centering
    \includegraphics[width=0.98\columnwidth]{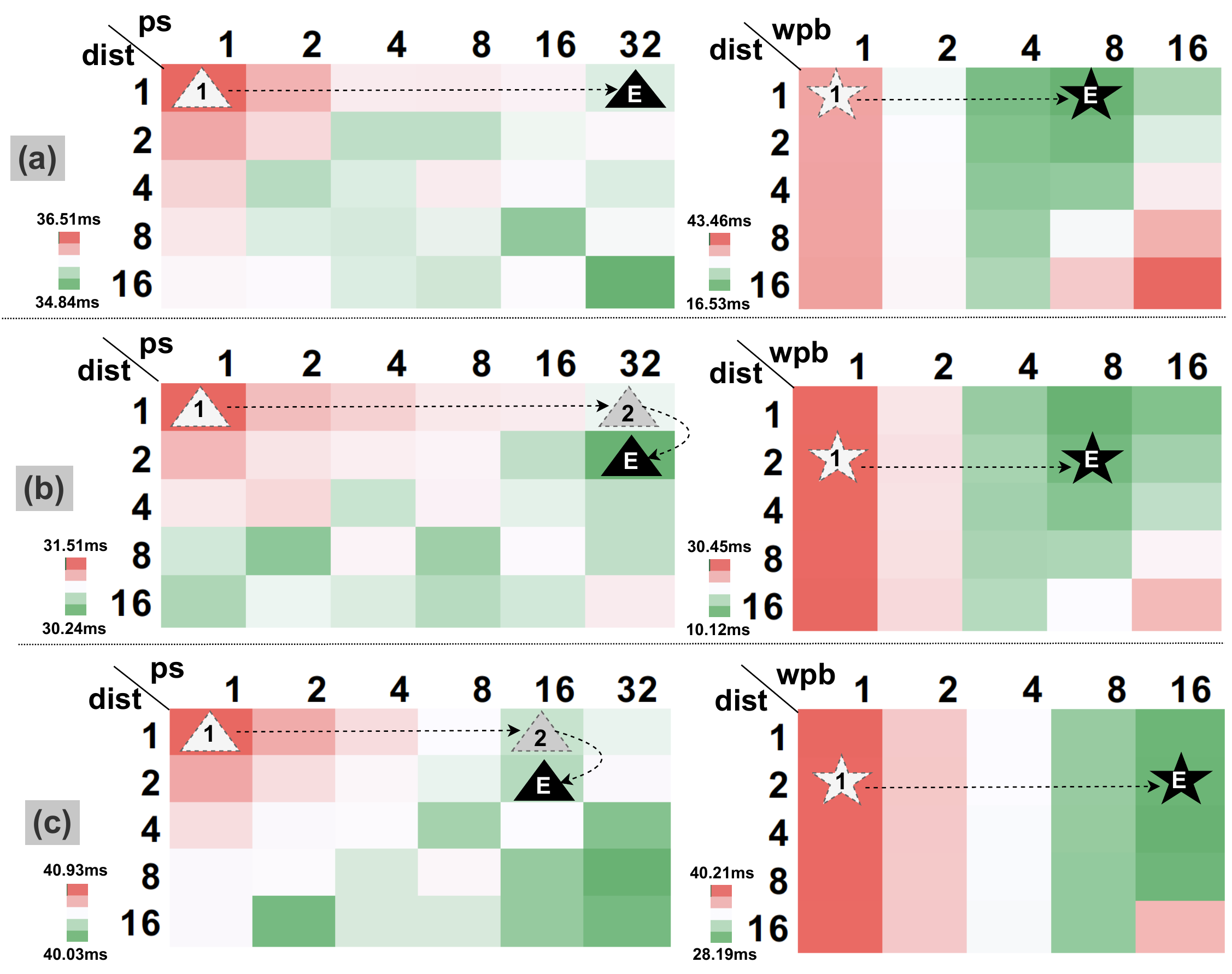}
    \caption{Parameter selection for three different settings. (a), (b), and (c) are for setting I, II, and III, respectively. Note that the left-side figures show the runtime latency for different combinations of $\mathit{ps}$ and $\mathit{dist}$, while the right-side figures show the latency for different combinations of $\mathit{wpb}$ and $\mathit{dist}$. The solid black triangle with ``E'' is the searched ``optimal'' combination for $\mathit{ps}$ and $\mathit{dist}$, while the black solid star with ``E'' is the searched ``optimal'' $\mathit{wpb}$ given $\mathit{dist}$ and $\mathit{ps}$.}
    \label{fig: Parameter Selection for Four Settings}
    \vspace{10pt}
\end{figure}

\textbf{Communication Primitives}
We adopt \Mname~with different NVSHMEM primitives at the \textit{thread}, \textit{warp}, and \textit{block} levels. We fix the number of GPUs to 2, the hidden dimension to 16, the neighbor partition size to 2, and the distance of workload interleaving to 2.
Figure~\ref{fig: Optimization Analysis}(c) shows that warp-level NVSHMEM primitives (\textit{e.g.}, \texttt{nvshmemx\_float\_warp\_get}) for remote accessing can bring the lowest latency. 
For thread-level NVSHMEM primitives (\textit{e.g.}, \texttt{nvshmem\_float\_get}), it would not coalesce the remote memory access to reduce unnecessary transactions. 
For the block-level NVSHMEM primitives (\textit{e.g.}, \texttt{nvshmemx\_float\_block\_get}), the higher overhead comes from collaborating a block of threads for remote access, since thread blocks (usually consisting of multiple warps) is larger than a single warp, thus, leading to higher synchronization and scheduling cost. This study also shows that our choice of warp-level primitives strikes a good balance between memory access efficiency and scheduling flexibility.

\textbf{Modeling and Optimization}
We further analyze the effectiveness of our lightweight analytical model for design space search. Specifically, three key parameters are studied, the size of neighbor partitioning (\textit{ps}), the interleaving distance (\textit{dist}), and the warps per block (\textit{wpb}).
We consider three different settings on a 2-layer GCN model: 
I: RDD on 4$\times$A100 as the basic setting.
II: RDD on 8$\times$A100 to demonstrate the adaptability toward the different numbers of GPUs.
III: RDD on 4$\times$V100~\cite{tesla-v100} to demonstrate the adaptability toward the different types of GPUs. We decompose searching results into two parts corresponding to the output of the \textbf{\textit{second}} and \textbf{\textit{third}} steps of the optimization discussed in $\S$\ref{sect: Modeling and Design Optimization}. 

Figure~\ref{fig: Parameter Selection for Four Settings} shows that our performance modeling and parameter selection strategy can pinpoint the low-latency design for the above three settings. The overall searching process only requires about 10 iterations to reach the final ``optimal'' settings. Note that here we show latency results for all possible settings for comparison. While in practice, we only need to traverse a small part of the whole design space (as indicated by the boxes touched by the dot lines).
By comparing the final optimal runtime configuration setting and the initial configuration, we can see that modeling and cross-iteration optimization can decrease the execution time by up to 68\%.
In the end-to-end GNN training (usually more than 100 iterations), such a latency saving would also be significant.

\subsection{Additional Study}
\label{sect: additional study}
\textbf{Accuracy-latency Tradeoff}
This study will analyze the accuracy-latency tradeoff between GNNs with sampling and full-graph (w/o sampling) on 8$\times$A100. Table~\ref{tbl: sampling vs non-sampling GNN} shows an evident node classification accuracy increase (2\% to 5\%) of GNN w/o sampling over GNN w/ sampling. The accuracy of sampling-based GNN would be affected by many factors (\textit{e.g.}, sampling rate at each GNN layer and graph structure). It is thus highly tricky to choose the ``optimal'' value for those factors. Here we follow the conventional way for GNN sampling~\cite{wang2019dgl}. The accuracy difference agrees with previous GNN algorithmic work~\cite{SageConv}. In many real-world applications (e.g, e-commerce), such an accuracy advantage of full-graph GNNs are be more preferred by users. Because even 1\% accuracy would make significant profit gains when deploying services at scale while the latency penalty is relatively minor. 
\begin{table}[t] \small
\centering
\caption{Accuracy-Latency of GNNs w/ and w/o sampling.}
\vspace{-5pt}
\scalebox{0.9}{
\begin{tabular}{l|>{\PreserveBackslash\raggedleft}m{1.8cm}|>{\PreserveBackslash\raggedleft}m{1.8cm}|>{\PreserveBackslash\raggedleft}m{2.3cm}}
\specialrule{.1em}{.05em}{.05em}
\textbf{Dataset} & \textbf{Accuracy w/ sampling} & \textbf{Accuracy w/o sampling} & \textbf{Latency (w/o \textit{vs.} w/ sampling)} \\ \hline\hline
RDD & 0.937 & 0.957 & 1.07$\times$ \\ \hline
PROT & 0.776 & 0.825 & 1.25$\times$ \\ 
\specialrule{.1em}{.05em}{.05em}
\end{tabular}}
\label{tbl: sampling vs non-sampling GNN}
\vspace{5pt}
\end{table}



\textbf{Generality to other applications} The design of \Mname~can be generalized to other similar applications.
We demonstrate the typical and popular deep-learning recommendation model (DLRM)~\cite{naumov2019deep, zhang2019deep, wang2022rec} that has been widely used in the industry. 
In multi-GPU DLRM, the large embedding tables are partitioned by rows and stored in different GPUs. The DLRM inputs (embedding access queries) will request embeddings from tables on different GPUs and then apply operations (e.g., elementwise addition or dot product) on those fetched embeddings. Such embedding lookup is highly sparse and irregular and dominates (> 80\% latency~\cite{zhang2019deep, gupta2020architectural}) the overall DLRM computation. 
We improve the mainstream DLRM system~\cite{naumov2019deep} with the design and optimizations of \Mname~to accelerate embedding lookup and element-wise addition and compare with the original system (which relies on NCCL)~\cite{naumov2019deep} under 4-GPU settings on the popular Criteo Kaggle~\cite{kaggle} dataset. 
\begin{table}[t] \small
\centering
\caption{DLRM~\cite{naumov2019deep} with \Mname~in Embedding Lookup.}
\vspace{-5pt}
\begin{tabular}{l|r|r }
\specialrule{.1em}{.05em}{.05em}
\textbf{Implementation} & DLRM~\cite{naumov2019deep} & DLRM (\Mname)  \\
\hline\hline
\textbf{Time (ms)} & 315.27 & 119.66  \\
\specialrule{.1em}{.05em}{.05em}
\end{tabular}
\label{tbl: pipeline on DLRM}
\vspace{6pt}
\end{table}
Table~\ref{tbl: pipeline on DLRM} shows that DLRM with \Mname~effectively reduces the lookup time (2.64$\times$). The fine-grained remote  access of \Mname~can reduce redundant inter-GPU traffic by using NCCL and offset the cost by massively parallel GPU-initiated communication.

\section{Discussion}
{\textbf{Deep Learning Pipelines: }}  {Despite the {popularity} of the pipeline concept in the conventional dense DL, the generalization of such a technique in sparse GNN computation is yet to be explored in-depth.
PiPAD~\cite{wang2023pipad} overlaps the communication (CPU-to-GPU) and processing (on GPUs) between adjacent graph partitions. Adopting this strategy, we will get designs as Figure~\ref{fig: Pipeline-aware Workload Management}(c), which would still suffer from pipeline bubbles due to workload imbalance.
vPipe~\cite{zhao2021v} dynamically assigns a DNN layer to certain pipeline stages during the runtime. It improves pipeline efficiency and GPU utilization for DNN models. However, adopting this approach in our fine-grained kernel pipeline would incur high overhead due to frequent workload reassignment and context switching.
In addition, the pipeline bubbles in dense DNN are predictable, input-agnostic, and can be reduced offline. However, the pipeline bubble for GNN can only be figured out at runtime due to input dependency. It, therefore, demands careful online workload balance and a pipeline schedule/mapping.}

{\textbf{Graph Partitioning Strategies: }} 
{Besides our current ID-based graph partitioning, {our designs/optimizations could also be extended to support other graph partitioning strategies from prior graph processing and GNN work}. There are several major categories.
{\textit{1) Locality-driven partitioning}} (e.g., Gemini~\cite{zhu2016gemini} and Rabbit order~\cite{rabbit-order}) minimizes the communication/synchronization cost in distributed graph processing/GNN computing. 
%
Such partition strategies are orthogonal to our current design optimization. Despite it will reduce the total size of communication, the communication pattern remains the same with irregular, sparse, and fine-grained data movements. Our \Mname~design can be modified to accommodate such reduced-communication cases through dynamic kernel re-configuration (e.g., fine-tuning the interleaving distance and warp-to-block mapping) to maximize communication and computation efficiency.
{\textit{2) Workload-driven partitioning}} (e.g., NeuGraph~\cite{ma2019neugraph} and CUBE~\cite{zhang2016exploring}) balances the irregular graph/GNN workload among different devices.
This type of strategy typically maintains multiple replicas of nodes and node properties on different devices and synchronizes partial results in replicas after local computation on each device.
Our current design be adapted to handle such cases by inserting device synchronization primitives (NVSHMEM collective communication primitives, such as \texttt{nvshmem\_float\_sum\_reduce}) for maintaining data consistency among different replicas.
{\textit{3) Learning-based partitioning}} (e.g., ROC~\cite{sysMLROC}) dynamically learns an ``optimal'' partitioning strategy that can maximize the computation performance. 
Our current design/optimization can also support this partitioning strategy by incorporating the overhead of NVSHMEM remote memory access in the runtime prediction model when optimizing partitioning strategies online.}

\section{Conclusion}
This paper presents MGG, a novel multi-GPU system design, and implementation to exploit the potential of leveraging GPU intra-kernel software pipeline for accelerating GNNs.
MGG consists of GNN-tailored pipeline construction and GPU-aware pipeline mapping to facilitate workload balancing and operation overlapping, and an intelligent runtime design to dynamically improve the GNN runtime performance.
Experiments show the advantages of MGG over state-of-the-art solutions and its generality towards other DL applications.


\section{Acknowledgment}
{We would like to appreciate the great help and support from OSDI shepherd and anonymous reviewers.}
This research was partially supported by the U.S. DOE Office of Science, Office of Advanced Scientific Computing Research, under award 66150: "CENATE - Center for Advanced Architecture Evaluation". The Pacific Northwest National Laboratory is operated by Battelle for the U.S. Department of Energy under Contract DE-AC05-76RL01830.
This work was also supported in part by NSF-2124039 and CloudBank~\cite{norman2021cloudbank}. 
In addition, we appreciate the generous help and support from Amazon Faculty Research Award 2021 for Professor Yufei Ding and NVIDIA Graduate Fellowship 2022-2023 for Yuke Wang.

\bibliographystyle{plain}
\bibliography{references}

\clearpage
\clearpage
\appendix
\section{Artifact Appendix}
\Mname~is a holistic runtime for exploiting intra-GPU-kernel communication-computation pipelining to accelerate multi-GPU GNNs.
\Mname~consists of two parts. 
The first part is the host-side CPU program. It is responsible for dataset loading, runtime configuration generation, and invoking the GPU-side program. 
The second part is the device-side GPU program, called kernels. It is responsible for the major computation and communication of the GNN model on sparse neighbor-aggregation across GPUs and dense node-update phase within each GPU.
\Mname~introduces GNN-tailored pipeline construction and GPU-aware pipeline mapping to facilitate workload balancing and operation overlapping.


\begin{itemize}
\itemsep0em 
     \item Code repository: \textbf{Github}\footnote{\url{https://github.com/YukeWang96/MGG-OSDI23-AE.git}} and \textbf{Zenodo}\footnote{\url{https://doi.org/10.5281/zenodo.7853945}}.
    \item \textbf{Hardware, OS \& Compiler}:
    \begin{itemize}
    \itemsep0em 
 \item NVIDIA DGX-A100 with dual AMD Rome 7742 processors (each with 64 cores, 2.25 GHz), 1TB host memory, and 8$\times$A100 GPUs (40 GB) connected via NVSwitch (600 GB/s). 
 \item Operating systems: Ubuntu 20.04+.
  \item Compilers: NVCC (v11.2), GCC (v7.5.0), 
  \item Libraries: CUDA (v11.2), OpenMPI (v4.1.1), NVSHMEM (v2.0.3), cuDNN (v8.2).
\item Datasets: SNAP~\cite{snapnets} and OGB~\cite{hu2020open}.
\end{itemize}
\end{itemize}

\subsection*{Environment Setup}
\subsubsection*{Step-1: Download libraries and datasets.}
\textbf{-- 1.1. Download libraries.}
\begin{lstlisting}[style=tt1, numbers=none]
wget storage.googleapis.com/mgg_data/local.tar.gz
tar -zxvf local.tar.gz
tar -zxvf local/nvshmem_src_2.0.3-0/build_cu112.tar.gz 
\end{lstlisting}

\noindent \textbf{-- 1.2. Download datasets and Setup Baselines.}
\begin{lstlisting}[style=tt1,numbers=none]
wget storage.googleapis.com/mgg_data/dataset.tar.gz
tar -zxvf dataset.tar.gz
cd dgl_pydirect_internal/
wget storage.googleapis.com/mgg_data/graphdata.tar.gz 
&& tar -zxvf graphdata.tar.gz 
&& rm graphdata.tar.gz
cd ..
gsutil cp -r gs://mgg_data/roc-new/ .
\end{lstlisting}

\noindent \textbf{-- 1.3. Launch Docker for MGG.} 
\begin{lstlisting}[style=tt1,numbers=none]
cd docker 
./launch.sh
\end{lstlisting}

\noindent \textbf{-- 1.4. Compile MGG implementations.}
\begin{lstlisting}[style=tt1,numbers=none]
mkdir build && cd build && cmake .. && cd ..
./0_mgg_build.sh
\end{lstlisting}

\subsection*{Step-2. Run Initial Tests.}
\vspace{3pt}
\noindent {Please try below Section-3.4 and Section-3.5.}

\subsection*{Step-3: Experiments.}
\textbf{-- 3.1. Compare with UVM (Fig.8a and Fig.8b).}
\begin{lstlisting}[style=tt1,numbers=none]
./0_run_MGG_UVM_4GPU_GCN.sh
./0_run_MGG_UVM_4GPU_GIN.sh
./0_run_MGG_UVM_8GPU_GCN.sh
./0_run_MGG_UVM_8GPU_GIN.sh
\end{lstlisting}
Results can be found at \texttt{Fig\_8\_UVM\_MGG\_4GPU\_GCN.csv, Fig\_8\_UVM\_MGG\_4GPU\_GIN.csv, Fig\_8\_UVM\_MGG\_8GPU\_GCN.csv, Fig\_8\_UVM\_MGG\_8GPU\_GIN.csv}

\vspace{3pt}
\noindent \textbf{-- 3.2. Compare with DGL (Fig.7a and Fig.7b)}.
\begin{lstlisting}[style=tt1,numbers=none]
cd dgl_pydirect_internal/
./launch_docker.sh
cd gcn/
./0_run_gcn.sh
cd ../gin/
./0_run_gin.sh
\end{lstlisting}
Results of DGL can be found at \texttt{1\_dgl\_gin.csv} and \texttt{1\_dgl\_gcn.csv}. MGG reference is in \texttt{MGG\_GCN\_8GPU.csv} and \texttt{MGG\_8GPU\_GIN.csv}.

\vspace{3pt}
\noindent \textbf{-- 3.3. Compare with ROC on 8xA100 (Fig.9)}.
\begin{lstlisting}[style=tt1,numbers=none]
cd roc-new/docker
./launch.sh
\end{lstlisting}
Results can be found at 
\texttt{Fig\_9\_ROC\_MGG\_8GPU\_GCN.csv, Fig\_9\_ROC\_MGG\_8GPU\_GIN.csv}.

\vspace{3pt}
\noindent \textbf{-- 3.4. Compare NP with w/o NP (Fig.10a)}.
\begin{lstlisting}[style=tt1,numbers=none]
python 2_MGG_NP.py
\end{lstlisting}

Note that the results can be found at \texttt{MGG\_NP\_study.csv}.

\vspace{3pt}
\noindent \textbf{-- 3.5. Compare WL with w/o WL (Fig.10b)}.
\begin{lstlisting}[style=tt1,numbers=none]
python 3_MGG_WL.py
\end{lstlisting}
Note that the results can be found at \texttt{MGG\_WL\_study.csv}.

\vspace{3pt}
\noindent \textbf{-- 3.6. Compare API (Fig.10c)}.
\begin{lstlisting}[style=tt1,numbers=none]
python 4_MGG_API.py
\end{lstlisting}
Note that the results can be found at \texttt{MGG\_API\_study.csv}.

\vspace{3pt}
\noindent \textbf{-- 3.7. Design Space Search (Fig.11a).}
\begin{lstlisting}[style=tt1,numbers=none]
python 5_MGG_DSE_4GPU.py
python 5_MGG_DSE_8GPU.py
\end{lstlisting}
Results can be found at 
\texttt{Reddit\_4xA100\_dist\_ps.csv, Reddit\_4xA100\_dist\_wpb.csv,
Reddit\_8xA100\_dist\_ps.csv, Reddit\_8xA100\_dist\_wpb.csv}.

\end{document}